\documentclass[useAMS,usenatbib]{mn2e}
\usepackage{graphicx}
\usepackage{epstopdf}
\usepackage{amsmath}
\title[]{Time dependence of advection dominated accretion flow around a rotating compact object }
\author[F.Habibi]{
  F.Habibi $^{1}$\thanks{f\_habibi@Birjand.ac.ir}
\\
$^{1}$Department of Physics, University of Birjand, Brjand, Iran\\
}

\date{}
\begin{document}
\pagerange{\pageref{firstpage}--\pageref{lastpage}} \pubyear{2012}

\maketitle \label{firstpage}

\begin{abstract}
Time evolution of advection-dominated accretion flow
(ADAF) around a rotating compact object is presented.
The equations of time-dependent of fluid including the Coriolis force along with the centrifugal and pressure gradient forces are derived.
 In this research, it is assumed that
angular momentum transport is due to viscous turbulence and the $\alpha$-prescription is used
for the kinematic coefficient of viscosity. Moreover, the general relativistic effects are neglected. In order to solve the equations, we have used a
self-similar solution.  The solutions show that the behaviour of the physical quantities in a
dynamical ADAF is different from that for a steady accretion flow. Our results indicate that the physical quantities  are dependent of rotation parameter which is defined as the ratio of the intrinsic angular velocity of the central body to
the angular velocity of disc.
Also,  the effect of rotation parameter on these quantities  is different for co and counter-rotating flows. The solution shows that by increasing the rotation parameter $a$,  inflow-outflow region approaches  the central object for co-rotating flow and  moves outward for counter-rotating flow. We find that when flow is fully advection dominated  $(f \to 1)$, the entire gas has positive Bernoulli function.  Also, we suggest that the Bernoulli function  becomes more positive when the effect of
rotation on the structure of disc decreases.
\end{abstract}

\begin{keywords}
 accretion , accretion disc, compact object
\end{keywords}

\section{INTRODUCTION }
\label{sec:intro}

Accretion processes are now believed to play a major role in
many astrophysical objects, from protostars to discs around
compact stars and active galactic nuclei. Such systems have
been studied at different levels depending on their physical
properties. The modern standard theory was formulated in
Shakura (1972), Novikov \& Thorne (1973) and Shakura \& Sunyaev (1973). It provided remarkably successful contributions to understanding quasars, X-ray binaries and active galactic nuclei. One of the basic assumptions of this model is that viscously generated internal
energy is radiated out immediately. However, there exist distinct branches of accretion discs in which this assumption is violated.

One type of accretion discs that has attracted a lot of attention is advection-dominated accretion flows (Known as ADAFs).
 ADAFs were introduced to astrophysics by Ichimaru (1977)
  and have been studied extensively at different contexts by many researchers during
the past three decades (e.g., Abramowicz et al. 1988, Narayan \& Yi 1995a, Chen et al. 1995, Kato et al. 1996, Kaburaki 2000, Shadmehri 2004,  Abbassi et al. 2008, Khajenabi 2015, Samadi et al. 2017,2019).
 In this kind of flows, the energy released through viscosity may
be transported (advected) in
the radial direction toward the central object or stored in the flow
as entropy rather than being radiated. Therefore, ADAFs can occur in two distinct physical regimes. Firstly
they can occur when the accreting gas density and consequently
its optical depth become higher due to very high mass accretion
rates  (Begelman 1978, Begelman
\& Meier 1982).  In this limit radiation can be trapped in the
in-falling gas. Secondly, ADAFs can occur when the
infalling gas has low density and low optical depth when the
mass accretion rates become very small (Ichimaru 1977, Rees
et al. 1982, Narayan \& Yi 1994, 1995a,b, Chen 1995).
 In this limit radiative time scale becomes
longer than the accretion time scale and as a result almost all the
internal energy can be lost to the compact star (Bhatt \& Prasanna 2000).

 Accretion disc solutions for optically thin branch of ADAFs derived originally by Narayan  \& Yi (1994, hereafter NY94).
NY94  proposed a self-similar solution for one-dimensional problem that only the radial structure of the disc is studied in a detailed way.  These solutions have provided good insight into the properties of accretion flows. One
of the important results of this solutions is that  advection dominated flows are susceptible to produce outflows.
This result was confirmed with more exact calculations in their next works (Narayan \& Yi  1995a,b).
They argued that  a necessary condition for outflows is that the Bernoulli function become positive.  On the other hands, the positive value for Bernoulli function that is defined as the sum of the kinetic energy,  the potential energy and the enthalpy of the accretion disc,
 has been interpreted as meaning that the fluid is unbound and would
generate a powerful outflow. However, it does not imply a lack of conservation of energy, but simply arise because viscous stresses transfer energy from small to large radii (NY94).
 So, a number of authors have extended this
work and obtained related solutions (e.g.,  Honma 1996, Kato \& Nakamura 1998 , Blandford
\& Begelman 1999 , Manmoto et al. 2000).
However, our understanding of
accretion discs properties could be improved by a discussion of time-dependent hydrodynamical models. Ogilvie (1999) extended the steady state self-similar solutions of NY94 to the time-dependent case.
 In this approach, by introducing a dimensionless self-similar variable, an initial set of nonlinear partial
differential equations (i.e. function of r and t) is reduced to a set of ordinary differential equations (i.e. function of self-similar variable).
These solutions provide some insight into the dynamics of
 accretion and avoid many of the strictures of the steady self-similar solution.
 Now, this method has a wide range of applications in the full set of equations of MHD in many research fields
of astrophysics and is  followed by several researchers. For example, Shadmehri (2004) obtained a set of self-similar solutions for quasi-spherical accretion flows. He studied the effect of self-gravity on the evolution of accretion discs.
 Also, Khesali \& Faghei (2009, hereafter KF09)  examines the self-similar evolution of a ADAF  in the presence of a toroidal magnetic field. They show that the behaviour of the physical quantities in a
dynamical ADAF is different from that for a steady accretion flow.
 Habibi et al.(2015a) achieved a self-similar solution for time evolution of an accreting magneto-fluid with finite conductivity. They also extended these solutions for a thick accretion disc in two-dimensions (i.e. variables are functions of r, $\theta$  and t) (Habibi et al 2015b).

The effect of rotation of the central object on the dynamic of accretion discs  have been studied by several authors in both Newtonian and relativistic limits (Bahaskaran et. al 1990, Igumenshchev \& Abramowicz 1999, Prasanna \& Mukhopadhyay 2003, hereafter PM03, Bhattacharya et al 2010, Shaghaghian 2011).
 In a general relativistic framework, employing the Kerr background geometry we can account for
the effect of central object rotation in dynamic of disc. On the other hand, the general relativistic predicts that a massive object would distort the space-time around itself. However, the space-time around a rotating compact object is not only bent due to mass, it is also twisted due to rotation. This effect is the well Known as Lense Thirring effect or inertial frame dragging. Indeed, the orbits about a rotating object experience a torque due to Lense Thirring effect, which cause the plane of orbit to precess(Bardeen \& Petterson 1975). For an accretion disc, if this torque be comparable with internal viscous forces, the result is that the inner regions of disc are forced to align with the spin of the central compact object.On the other hands, Lense Thirring effect causes the
angular velocity of the gravitating source be proportional to  the
angular velocity of disc(PM03).
The frame dragging effect can be considered equivalent with the Coriolis type field in Newtonian physics (Abramowicz et al 1993). Therefore, if we want to create a similar scenario for an accretion disc around a rotating compact object in Newtonian limit , the only way would be writing of motion equations in a rotating frame that include the Coriolis force and Centrifugal force. PM03 studied the effect of Coriolis force on advection dominated accretion flow in a rotating frame.
They could solve the height integrated set of equations describing the steady state while assumed  ratio of the intrinsic angular velocity of the central body to the rotating velocity of disc is a constant value.
 The result of their work indicated that a simple coupling between the angular velocity of the central body and the angular velocity of the fluid element can show the possibility of energy transfer both inwards and outwards depending on the effective angular velocity of the fluid flow.
Here, we also look for possible influence of Coriolis type terms
on the time evolution of accretion discs around rotating compact objects, particularly for
the type wherein the disc is in contact with the central body.
 Therefore, we will consider a set of time-dependent equations governing the motion of an advection dominated accretion flow. So, we will solve the fluid equations for an accreting gas that is
self-similar over time.
 Although some effort have been made to solve time-dependent equations of accretion discs by self-similar technique (e.g., KF09, Habibi et al. 2015), all of them considered an accretion disc around non-rotating central object. In the present work, we aim to obtain more realistic view of this problem by adding the influence of the central object's rotation.
 There are two accreting ways, which depend on the rotating direction
of accretion flow relative to that of central object: (i) co-rotating that accretion flow is in the same direction as a spinning compact object; (ii) counter-rotating that accretion flow is in the opposite direction to a spinning compact object.
 We also will show that time evolution of  the co and
counter-rotating accretion flows onto the central compact object can be distinguished.
 This paper is organized as
follows. In Section 2, we define the general problem of constructing
a model for  accretion flow around a rotating compact object.
In Section 3, we use the self-similar method to solve the integrated
equations that govern the dynamical behaviour of the accreting gas. Finally,
we present a summary of the model in Section 4.

\section[]{BASIC EQUATION}
In this section, we derive the basic equations that describe
time evolution of accretion flow around a rotating compact object. To begin with, we consider the following standard assumptions:\\
(i) We use the spherical coordinates $(r, \theta , \varphi)$ centred on the accreting object and neglect all terms with any $\theta$ and $\varphi$
dependence, hence all physical quantities will be expressed in
terms of spherical radius $r$ and time $t$.\\
(ii) For simplicity, the general relativistic effects
have been neglected.\\
(iii)The gravitational field is produced by a slowly rotating compact object and
the gravitational force on a fluid element is characterized
by the Newtonian potential,$\Phi = \frac{G M}{r^2}$, with
$G$ representing the gravitational constant, $M$ standing for the
mass of the central compact object.\\
(iv)  $r\phi$-component of the viscous stress tensor is important and other components are neglected.\\

On the basis of these assumptions and simplifications, we can formulate
the basic equations governing the disc evolution as follow: \\
the continuity equation,
\begin{equation}
\frac{\partial \rho}{\partial t}+\frac{1}{r^2}\frac{\partial}{\partial r}(r^2 \rho v_r)=0,\label{1}
\end{equation}
the equations of motion
\begin{equation}
\frac{\partial v_r}{\partial t}+v_r \frac{\partial v_r}{\partial r} +\frac{1}{\rho} \frac{\partial p}{\partial r} +\frac{G M}{r^2}=2\omega \Omega r +\omega^2 r + r \Omega^2,\label{2}
\end{equation}
\begin{equation}
\frac{\partial}{\partial t}( r^2\Omega)+v_r \frac{\partial}{\partial r}(r^2\Omega)=\frac{1}{\rho r^2}\frac{\partial}{\partial r}\bigg[\nu \rho r^4 \frac{\partial \Omega}{\partial r}\bigg]-2\omega r v_r, \label{3}
\end{equation}
and the energy equation
\begin{equation}
\frac{1}{\gamma-1}\bigg[\frac{\partial p}{\partial t}+v_r \frac{\partial p}{\partial r}\bigg]+\frac{\gamma}{\gamma-1}\frac{p}{r^2}\frac{\partial}{\partial r}(r^2 v_r)= f \nu \rho r^2 (\frac{\partial \Omega}{\partial r})^2.\label{4}
\end{equation}
wherein $\omega$ is the intrinsic angular velocity of the central body, $v_r$ radial velocity and $\Omega(=v_\varphi /r)$ angular speed of the fluid element, $\rho$ density, $p$ the gas pressure and $\gamma$ is the ratio of specific heats.
The third term on the right hand side of the equation (\ref{2}) and the first term on the right hand side of the equation (\ref{3}) are components of Coriol force in unit of mass.
We take the kinematic coefficient of shear viscosity to be
 \begin{equation}
 \nu=\alpha\frac{p}{\rho \Omega_k}.\label{21}
 \end{equation}
where $\Omega_k$ is Keplerian angular velocity defined by
$\Omega_k=\sqrt{\frac{G M}{r^3}}$ (Shakura \&
Sunyaev 1973). The parameters $\alpha$ and $\gamma$ are constant. Also, the constant parameter $f$ shows the degree to which the flow is advection-dominated. In the extreme
limit of no radiative cooling, we have $f=1$, while in the opposite limit of very efficient cooling, $f=0$ ( NY94).

There are five unknown variables: $v_r, \omega, p, \rho$ and $\Omega$ in
four equations (\ref{1})-(\ref{4}), so we need an extra equation. This equation is a relation between the angular velocity of  the central object $\omega$ and the angular velocity of disc $\Omega$.
As  mentioned in the previous section, the angular velocity of  the central object can be proportional with the angular velocity of fluid element.
 Here following PM03 we adopt
 $\omega=a \Omega$,
where $a$ is rotation parameter. For simplicity,  $a$  is assumed to be constant. Therefore, we can rewrite equations (\ref{2}) and (\ref{3}) as:
\begin{equation}
\frac{\partial v_r}{\partial t}+v_r \frac{\partial v_r}{\partial r} +\frac{1}{\rho} \frac{\partial p}{\partial r} +\frac{G M}{r^2}= r (a+1 )^2 \Omega^2,\label{21}
\end{equation}
$$
\frac{\partial}{\partial t}( r^2\Omega)+v_r \frac{\partial}{\partial r}(r^2\Omega)=\frac{1}{\rho r^2}\frac{\partial}{\partial r}\bigg[\alpha \frac{p}{\sqrt{GM}} r^{11/2} \frac{\partial \Omega}{\partial r}\bigg]
$$
\begin{equation}
-2 a r\Omega v_r. \label{31}
\end{equation}
Thus, the basic equations governing the time evolution of a advection dominated accretion flow are represented by equations (\ref{1}), (\ref{4}),(\ref{21}) and(\ref{31}).
It is clear that the above equations are nonlinear and  we are not able to solve them analytically. In these circumstances,  self-similar method is very useful and widely adopted in the astrophysical literature. Generally self-similar solutions are divided into two
chief categories: temporal self-similar answers and spacial
self- similar answers.
 In next section, we will solve the basic equations of system with the aid of temporal self-similar technique.

\section{Self-similar solution}
\subsection{Analysis}
Because the equations of the system
depend on time, we can search for answers that describe
the temporal change of physical quantities in a way that
the change of each quantity at any instant of time is similar to the others.
Such answers are known as temporal self-similar solutions.
In this method, we should convert the basic equations in real space to a set of non-dimensional equations in similarity space.
To this aim, a combination of radius $r$ and time $t$ is introduced as dimensionless similarity variable as:
\begin{equation}
\zeta=\frac{r}{\delta t^n},\label{7}
\end{equation}
wherein $\delta$ is a constant that is used to make $\zeta$ dimensionless.
So, all physical quantities
are assumed as unknown powers of $t$, unknown functions of $\zeta$ and constant dimensional coefficients:
\begin{equation}
\rho(r,t)= R_0 t^{\epsilon_\rho} R(\zeta),\label{71}
\end{equation}
\begin{equation}
p(r,t)= P_0 t^{\epsilon_p} P(\zeta),\label{72}
\end{equation}
\begin{equation}
v_r(r,t)= V_0 t^{\epsilon_v}V(\zeta),\label{73}
\end{equation}
\begin{equation}
\Omega(r,t)= W_0 t^{\epsilon_w} W(\zeta).\label{8}
\end{equation}
Also, considering the transformation
$(r,t)\rightarrow(\zeta,t)$, derivatives will turn into
\[
\frac{\partial}{\partial r}\rightarrow \frac{\zeta}{r}
\frac{\partial}{\partial \zeta},
 \qquad \frac{\partial}{\partial t}\rightarrow
\frac{\partial}{\partial t}-n \frac{\zeta}{t} \frac{\partial}{\partial
\zeta}.
\]
Now, we can put these
answers into equations (\ref{1}), (\ref{4}),(\ref{21}) and (\ref{31})
to determine
the unknown powers $t$ and constant dimensional coefficients in a way that satisfy the equations.
we obtain the constant exponents as
\begin{equation}
n =\frac{ 2}{3}, \epsilon_v = -\frac{1}{3}, \epsilon_w = -1,
\epsilon_p = \epsilon_\rho -n,   \label{self8}
\end{equation}
and the dimensional coefficients as
\begin{equation}
 V_0  = \sqrt{\frac{GM}{\delta}}, W_0=\sqrt{\frac{GM}{\delta^3}}, P_0 =\frac{GM}{\delta} R_0, \delta=(GM)^{1/3}  .\label{81}
\end{equation}
The above results show that the radius of the flow increases proportionally to $t^{2/3}$.
The time dependent behaviour of the radial velocity is proportional to $t^{-1/3}$, namely,
it slow down as time goes by.  Also, the angular speed
scales with time as $t^{-1}$.
 Moreover, based on these results the time behaviour of gas pressure is dependent on the time behaviour of density.
 These results are in agreement with the previous
similar studies (Habibi et al 2015, KF09). 
 To determine the behavior of density and pressure with time we need another relation.
 we introduce the mass accretion rate $\dot{M}$ as:
\begin{equation}
\dot{M}=-4\pi r^2 \rho v_r.\label{11}
\end{equation}
Similarly to other physical quantities, the mass accretion rate can
consider in similarity space as
\begin{equation}
\dot{M}(r,t)= M_0 t^{\epsilon_M} M(\zeta).\label{12}
\end{equation}
Under transformations (\ref{7}), (\ref{71}) and (\ref{73}), the mass accretion rate can also write in following form:
\begin{equation}
\dot{M}(r,t)
= \bigg(\delta^3 R0 \bigg) t^{2n+\epsilon_\rho-1/3} \bigg(-4\pi \zeta^2 R(\zeta)V(\zeta)\bigg).\label{13}
\end{equation}
So, equality of equations (\ref{12}) and (\ref{13}) eventuate:
\begin{equation}
M_0=\delta^3 R0 , \epsilon_M=2n+\epsilon_\rho-1/3, M(\zeta)=-4\pi \zeta^2 R(\zeta)V(\zeta).
\end{equation}
Here following KF09, we consider a set of solutions that mass accretion rate is independent of time, thus we have
\begin{equation}
R_0=\frac{M_0}{G M}, \epsilon_\rho=-1,
\end{equation}
and
\begin{equation}
P_0=M_0(G M)^{-1/3}, \epsilon_p=-\frac{5}{3}.
\end{equation}
With self-similar
solutions (\ref{7})-(\ref{81}), the basic equations (\ref{1}), (\ref{4}), (\ref{21}) and (\ref{31}) are reduced to a
set of comparatively simple ordinary differential equations as follow:
\begin{equation}
-R+\bigg(V-\frac{2}{3}\zeta\bigg)\frac{d R}{d\zeta}+\frac{R}{\zeta^2}\frac{d}{d\zeta}(\zeta^2 V)=0,\label{34}
\end{equation}
\begin{equation}
-\frac{1}{3}V +\bigg(V-\frac{2}{3}\zeta\bigg)\frac{dV}{d\zeta}+\frac{1}{R}\frac{dP}{d\zeta}+\frac{1}{\zeta^2}= (a+1)^2\zeta W^2,\label{35}
\end{equation}
$$
\frac{1}{3}(\zeta^2 W)+\bigg(V-\frac{2}{3}\zeta\bigg)\frac{d}{d\zeta}(\zeta^2 W)= \frac{\alpha}{R \zeta^2}\frac{d}{d\zeta}\bigg[P \zeta^{11/2}\frac{dW}{d\zeta}\bigg]
$$
\begin{equation}
-2a \zeta W V,\label{36}
\end{equation}
$$
\frac{1}{\gamma-1}\bigg[-\frac{5}{3}P+\bigg(V-\frac{2}{3}\zeta\bigg)\frac{dP}{d\zeta}\bigg]+\frac{\gamma}{\gamma-1}\frac{P}{\zeta^2}\frac{d}{d\zeta}(\zeta^2V)
$$
\begin{equation}
=\alpha f P \zeta^{7/2}\big(\frac{dW}{d\zeta}\big)^2,\label{37}
\end{equation}
 Once the constant parameters are selected, the set of above equations can be numerically
solved with proper boundary conditions. For this purpose and as first step, we will follow the asymptotic behavior of the physical quantities  in next subsection.

\begin{figure*}
\centering
\includegraphics[width=180mm]{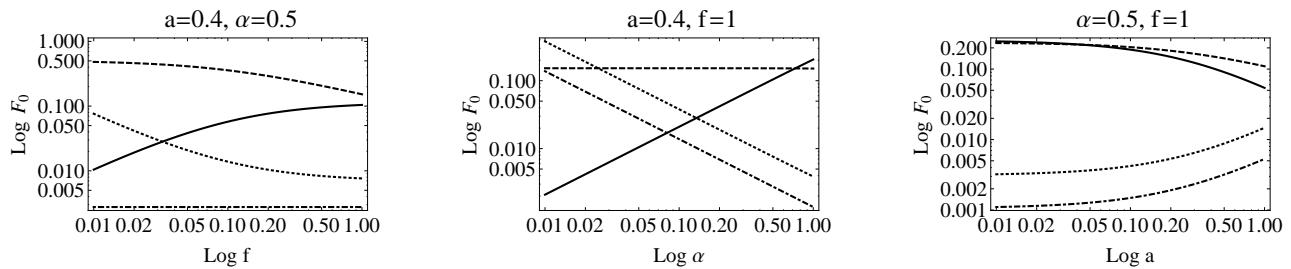}
\caption{Numerical coefficients $R_0$ (dotted lines), $V_0$ (solid lines), $W_0$ (dashed lines) and $P_0$ (dot-dashed lines) as functions of the advection parameter $f$, the viscous parameter $\alpha$ and rotation parameter $a$.
The ratio of specific heats is set to be $\gamma = 1.5$ and the inner mass accretion rate is $\dot{m}_ {in}$ = 0.01.}\label{figasy}
\end{figure*}

Before solving the above equations numerically, it is intended to study the possibility of outflow  in accretion flow. Therefore, we introduce Bernoulli function as
\[
Be =  \frac{1}{2}V^2+\Phi+W,
\]
where $W$ is the specific enthalpy, $V$ is the velocity (all three
components included) and $\Phi$ is the gravitational potential.
The positive value of Bernoulli function indicates the ability of accretion matter to reach infinity with a net positive
kinetic energy.
The Bernoulli function of advection dominated
flow can rewrite as:
\begin{equation}
Be=\frac{1}{2}(v_r^2+r^2\Omega^2)-\frac{GM}{r}+\frac{\gamma}{\gamma-1}\frac{p}{\rho}.
\end{equation}
By using transformations of equations (\ref{7})-(\ref{8}) and self- similar
solutions (\ref{self8})-(\ref{81}), Bernoulli function can be expressed in the following form
\[
Be(r,t)=\frac{GM}{\delta} t^{-\frac{2}{3}}\bigg\{\frac{1}{2}\big[V(\zeta)^2+\zeta^2 W(\zeta)^2\big]-\frac{1}{\zeta}+\frac{\gamma}{\gamma-1}\frac{P(\zeta)}{R(\zeta}\bigg\}
\]
\begin{equation}
=v_0(t)^2 b(\zeta).
\end{equation}
Here,  $b(\zeta)=\frac{Be}{v_0(t)^2}$ is  dimensionless Bernoulli parameter in a temporal self-similar flow and $v_0(t)= \sqrt{\frac{GM}{\delta}} t^{-\frac{1}{3}}$ (Ogilvie 1999).
We will follow the behavior of Bernoulli parameter for self-similar solutions obtained in this paper in the next
subsection.


\begin{figure*}
\centering
\includegraphics[width=160mm]{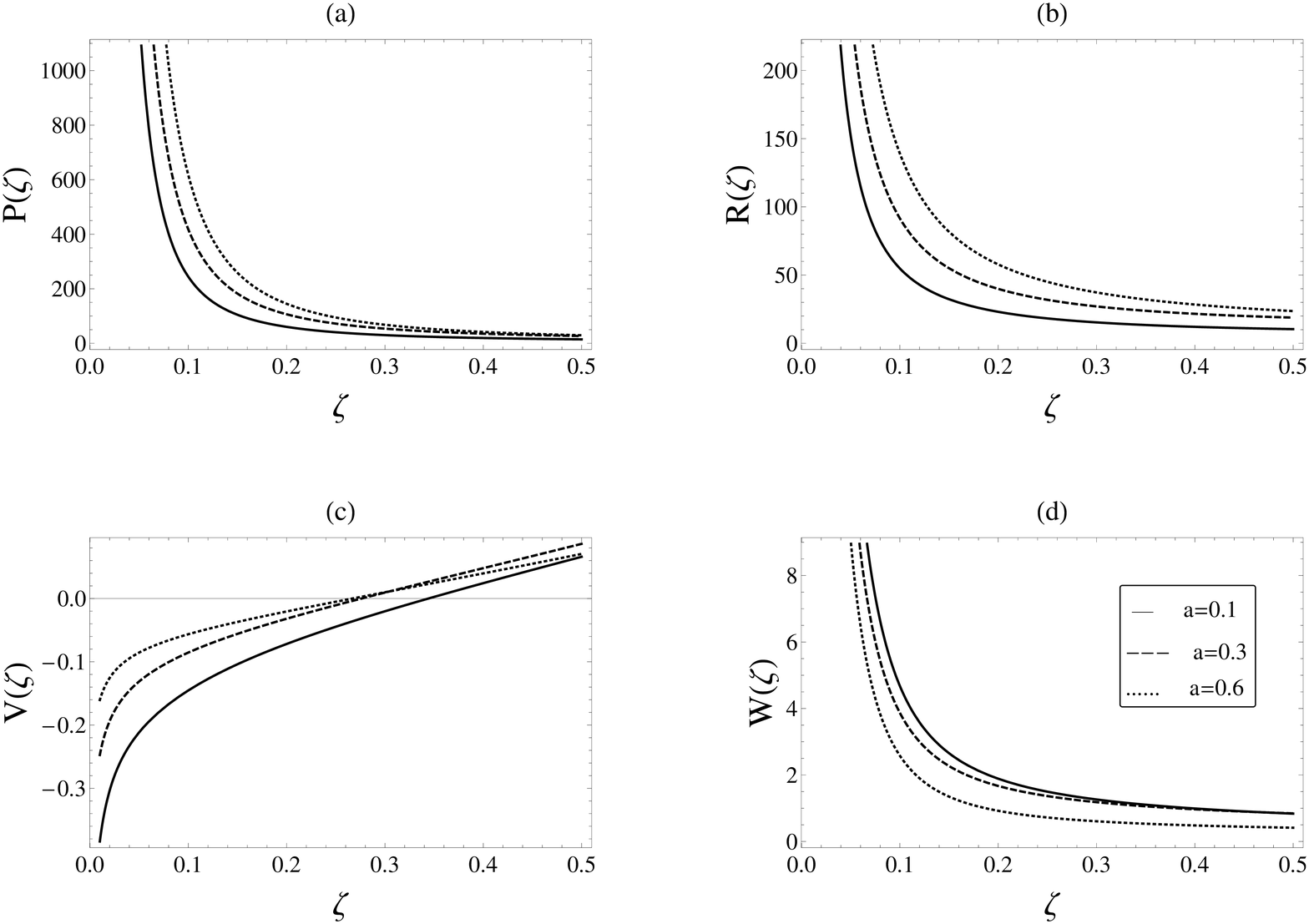}
\caption{Time-dependent self-similar solutions for a co-rotating flow as a function $\zeta$ for different values of the rotation parameter $a$. Constant parameters are: $\alpha=0.3$, $f=1$ and $\gamma=1.5$}\label{figa1}
\end{figure*}

\subsection{Asymptotic behavior }\label{S5}
The asymptotic behaviour of the equations (\ref{34}) - (\ref{37}) at limit $\zeta \to 0$ can be obtained in the following form:
\begin{equation}
R(\zeta) \sim R_0 \zeta^{-3/2},
\end{equation}
\begin{equation}
P(\zeta) \sim P_0 \zeta^{-5/2},
\end{equation}
\begin{equation}
V(\zeta) \sim V_0 \zeta^{-1/2},
\end{equation}
\begin{equation}
W(\zeta) \sim W_0 \zeta^{-3/2},
\end{equation}
in which
 \begin{equation}
R_0=\frac{3 \alpha}{4 \pi}\frac{\dot{m}_{in}}{A g},
\end{equation}
\begin{equation}
P_0=\frac{\dot{m}_{in}}{6 \pi \alpha}(4a+1),
\end{equation}
\begin{equation}
V_0=-\frac{A g}{3 \alpha},
\end{equation}
\begin{equation}
W_0=\frac{\sqrt{\epsilon A g}}{3 \alpha},
\end{equation}
where
\begin{equation}
\epsilon =\frac{5/3-\gamma}{(\gamma-1)f},
\end{equation}
\begin{equation}
A=2 n \epsilon +5 (1+4a),
\end{equation}
\begin{equation}
n=(1+a)^2,
\end{equation}
\begin{equation}
g=-1+\sqrt{1+\frac{18 \alpha^2}{A^2}},
\end{equation}
and $\dot{m}_{in}$  is the value of mass accretion rate at a point near to the centre.
Our asymptotic solutions show the same radial dependence with the steady solutions of PM03. This is a logical consequence because $\zeta \to 0 $ can be corresponded to $t \to \infty$ for fixed $r$. It means that system can be expected to approach a steady state (Ogilvie 1999).
The results achieved for asymptotic behaviour of physical quantities show that the physical quantities of accretion flow are
sensitive to parameters $\alpha, \gamma, f$ and  $a$. The effect of the viscous parameter $\alpha$, the advection parameter $f$ and the rotation parameter $a$ on the asymptotic solutions
  are plotted in figure \ref{figasy}.

The left panel of figure \ref{figasy} shows the variations of  numerical coefficients of physical quantities for some values
of $f$. We see that by adding $f$ the numerical coefficients of density $R_0$ and angular velocity $W_0$ decrease but the numerical coefficient of the radial infall velocity $V_0$ increases. Moreover, the numerical coefficient of pressure do not have dependence on the value of $f$ very much. These results are similar to the results of KF09.
The middle panel of figure \ref{figasy} displays numerical coefficients of physical quantities with respect to $\alpha$.
 It clearly shows that when the viscous parameter becomes larger,  the numerical coefficients of density, pressure and angular velocity reduce.
That's because we increase the viscous torque by increasing parameter $\alpha$. However, the numerical coefficient of the radial velocity $V_0$
rises with increasing $\alpha$ which is similar to the result of  KF09.
In the right panel of figure \ref{figasy},  we can see how the rotation  parameter $a$ affects on numerical coefficients of physical quantities.
 It shows that for higher values of $a$, both the numerical coefficients of  the radial velocity and angular velocity slow down but the numerical coefficients of the density and pressure grow. The behavior of solutions with respect to $a$ is qualitatively consistent with the results of PM03.

\begin{figure*}
\centering
\includegraphics[width=160mm]{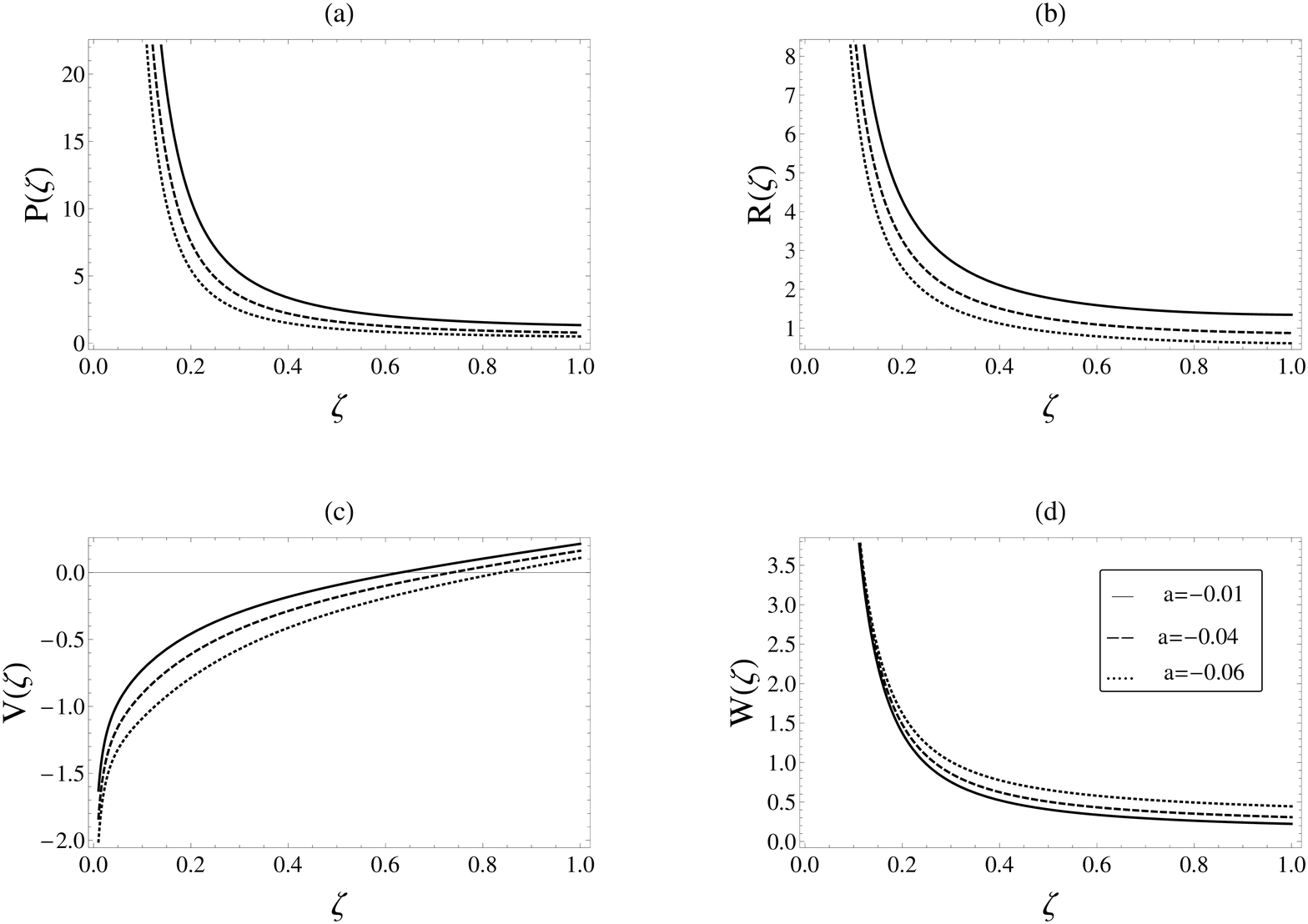}
\caption{Time-dependent self-similar solutions for a counter-rotating as a function $\zeta$ for different values of the rotation parameter $a$. Constant parameters are: $\alpha=0.3$, $f=1$ and $\gamma=1.5$}\label{figa2}
\end{figure*}

\subsection{Numerical Solutions}

Using the above asymptotic solutions, the equations (\ref{34})-(\ref{37}) can be numerically integrated
from a point near to the centre. Examples of such solutions
are presented in Fig \ref{figa1} and Fig \ref{figa2} for the cases co-rotating
($a> 0$)  and counter-rotating ($a < 0$), respectively.
 Here, we have focused on the
effect of rotation on the dynamic of disc.
Therefore,  we have plotted physical quantities with respect to $\zeta$ for fixed values of $\alpha$, $\gamma$ and $f$.  The general behavior of the various physical quantities are similar for all values
of $a$, i.e. these quantities gradually grow inward.
 However, as is illustrated by Fig \ref{figa1} for the co-rotating case,
 a rise in the rotation parameter $a$, slows down both velocity components and leads to an increase in the density and the gas pressure.
 In figure \ref{figa2}, we can see how the physical quantities for a counter-rotating flow change with variation of the rotation parameter $a$.
 Unlike Fig \ref{figa1}, for higher absolute values of $a$ both radial and  angular velocities increase but the density and the pressure of gas become smaller in a given $\zeta$. As a result, the effect of the rotation  parameter $a$ on the physical quantities for the co-rotating flow is exactly contrary to counter-rotating flow.
Moreover, we notice that the  effect of rotation is more remarkable in innermost regions of the disc.
Our results are qualitatively in agreement with the previous
similar studies (PM03, Bhaskaran et al 1990).
Furthermore, according to the sign of radial velocity in the panel (c) of figure \ref{figa1}, we can define
two regions. The first one is called inflow region, which the radial velocity has a negative
value and the accretion material moves toward the central object. The next area, which the radial velocity has a positive
value and means that the flow moves away from the central object, is called outflow region.
There is  a particular point where
the radial velocity becomes  equal to zero. This point distinguishes the inflow and outflow regions.
It can be called as the stagnation point.
From the Fig \ref{figa1}(c) it is clear that the rotation  parameter $a$ have a significant effect on inflow and outflow regions.
By adding the rotation parameter $a$, the stagnation point approaches the central object and therefore outflow occur in a position nearer to the center.
Similar to Fig \ref{figa1}(c), we can see both inflow and outflow regions for a counter-rotating flow in Fig \ref{figa2}(c).
 As this figure represents the stagnation point  moves outward for a larger value of $a$. Thus, for counter-rotating case, the inflow area increases with increasing $a$ but outflow region decrease. As a result, the effect of the rotation parameter $a$ on the stagnation point for counter-rotating flow is entirely opposite to co-rotating flow.
Our results are qualitatively consistent with the previous results (Bhattacharya et al 2010, Igumenshchev \& Abramowicz 1999).

\begin{figure*}
\centering
\includegraphics[width=160mm]{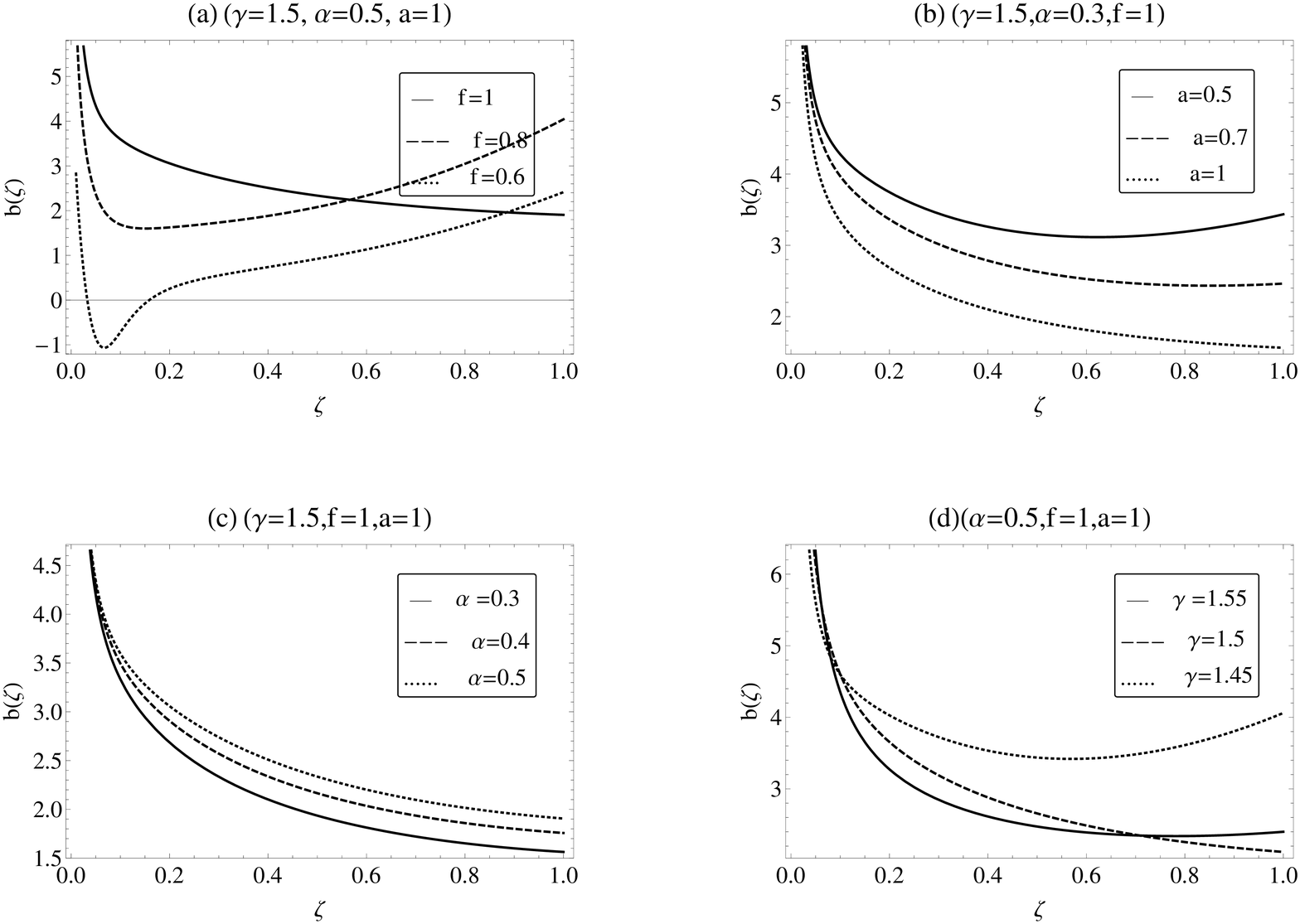}
\caption{Dimensionless Bernoulli parameter for a co-rotating flow with respect to  $\zeta$  and several values of $f$, $a$, $\alpha$ and $\gamma$}\label{figb1}
\end{figure*}

Figure \ref{figb1} shows the Bernoulli parameter $b$ as a function of $\zeta$ for different values of parameters $\alpha, \gamma, f$ and $a$ in a co-rotating flow. These plots display that advection dominated flows can have positive $b$. As is emphasized by NY94, this is a unique feature for ADAFs.
In Fig \ref{figb1}(a), the effect of advection parameter $f$ on the dimensionless Bernoulli parameter is showed  for the fixed values of $\alpha=0.5, \gamma=1.5$ and $a=1$. As this plot clearly shows for flows that are highly advection dominated, the entire gas has positive Bernoulli parameter.  Indeed, by increasing advection parameter $f$ the viscous energy flux decreases, a larger fraction of the energy is retained by the gas and therefore, $b$ becomes larger (Narayan \& Yi 1995a).
The dimensionless Bernoulli parameter $b$ is demonstrated for a fully advection dominated flow ($f=1$) with $\gamma=1.5,\alpha=0.3$ and several values of $a$ in Fig \ref{figb1}(b). As a result, the Bernoulli parameter becomes more positive when we employ the values of smaller for the rotation parameter. This is the effect of Coriolis force.
From the bottom left panel of Fig \ref{figb1}, we see that the Bernoulli parameter for a fully advection dominated flow ($f=1$) grows as the effect of viscous parameter becomes  higher.  Moreover, this effect is important at larger radii.
The effect of $\gamma$ on the Bernoulli parameter $b$ is specified in the bottom right panel of Fig \ref{figb1}.
According to this plot, with larger
values $\gamma$, the Bernoulli parameter  becomes smaller for a fully advection dominated flow ($f=1$). This is resulting from decreasing contribution of enthalpy in the Bernoulli function.
.

\begin{figure*}
\centering
\includegraphics[width=160mm]{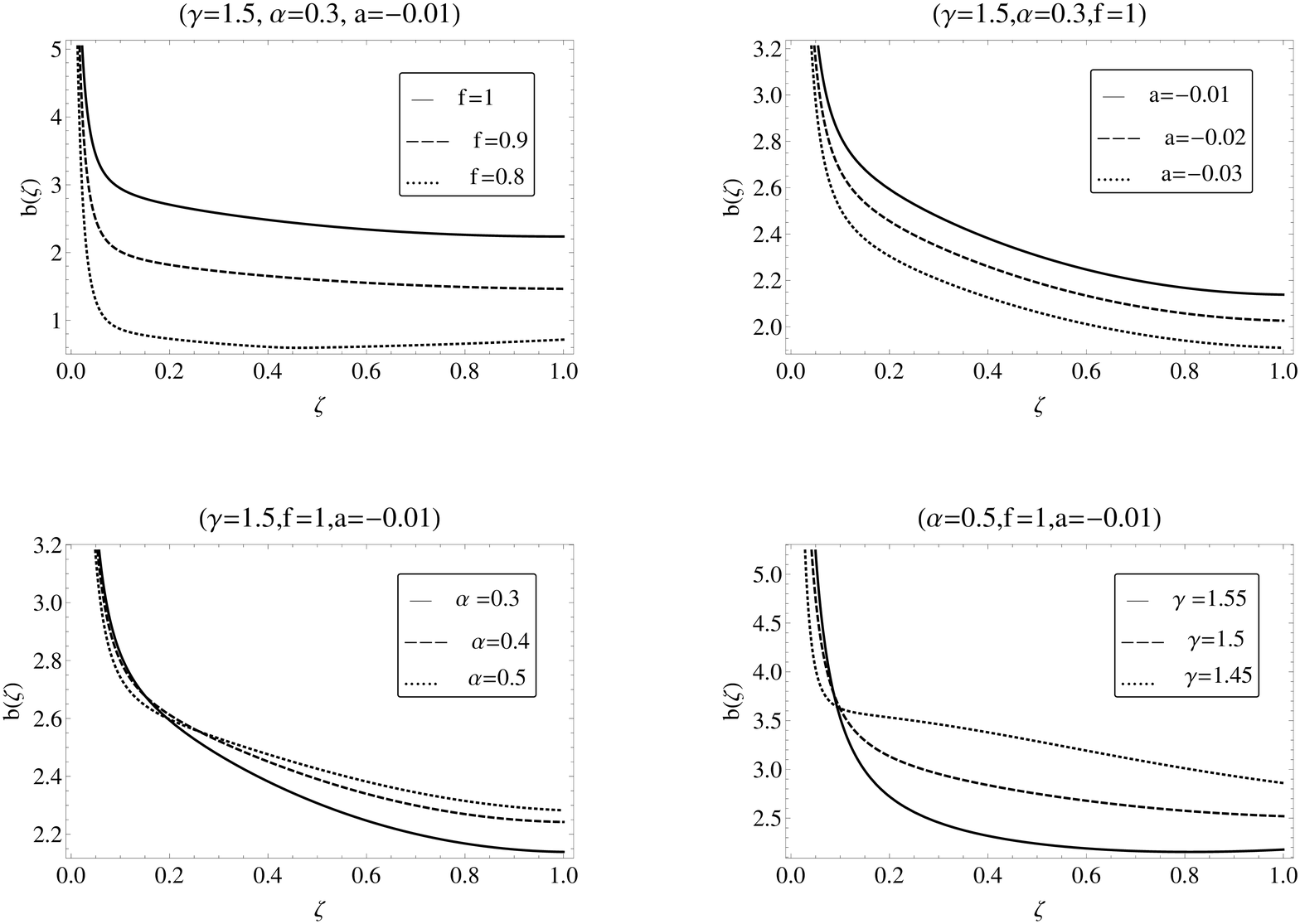}
\caption{Dimensionless Bernoulli parameter for a counter-rotating flow with respect to  $\zeta$  and several values of $f$, $a$, $\alpha$ and $\gamma$}\label{figb2}
\end{figure*}

Fig \ref{figb2} is plotted similarly to Fig \ref{figb1} but for a counter-rotating flow. Here  as we expected, the Bernoulli parameter  behaves as the same as co-rotating flow, i.e. by increasing the values of  $\gamma$ and $a$ the bernoulli parameter reduces for a given $\zeta$  while it becomes more positive by adding $f$ and $\alpha$.
This is reasonable
because energy do not depends on the rotating direction.

\section{Conclusions}
In this paper, the equations of a time-dependent advection dominated accretion flow around a rotating compact object have been
solved by semi-analytical similarity methods. We have restricted ourselves to fluid flows
in which general relativistic effects are negligible. We also have used the usual $\alpha$ -prescription for the viscosity.
 The flow is not able
to radiate efficiently, so the energy equation is used. We have driven the equations of motion in a rotating frame that include the Coriolis force and
Centrifugal force.  A direct relationship between the angular velocity of the central object and the angular velocity of
fluid element as $\omega=a\Omega$ is employed. Here, $a$ is called the rotation parameter.
The introduction of the Coriolis force
 and a direct coupling between of
the angular velocity of the central object and the angular velocity of
fluid element seem to bring in some features which could be helpful for understanding
the physical behavior of disc.

In self-similar space, the radius of the flow increases proportionally to $t^{2/3}$. The
time dependent behavior of the radial velocity is proportional to $t^{-1/3}$ and  the angular speed is scaled with time as $t^{-1}$.
But the time dependent behavior of pressure of gas is dependent on the time behaviour
of density. We  have restricted ourselves to fluid flows
in which mass accretion  rate  is constant with time. Therefore, we have obtained time behavior of the density and pressure as $t^{-1}$ and  $t^{-5/2}$, respectively.

The radial dependence of the calculated physical quantities in this approach is different from that for a steady
self-similar solution ( Prasanna \& Mukhopadhyay 2003) . Also, our results show that the physical quantities of accretion flow are sensitive to $\alpha$ parameter, the advection parameter $f$ and the rotation parameter $a$.
We have focused our work to study the effect of spin of black hole and role of it in the dynamics of accretion disc. Therefore, the behavior of physical quantities studied for different values of the rotation parameter $a$ with  $\alpha, \gamma$ and $f$ fixed.
Our results show that the effect of rotation parameter is different for co and counter rotating flows.
In a co-rotating flow,  the gas pressure and density rise by increasing $a$ but the radial velocity and the angular velocity diminish.
On the contrary, for a counter-rotating flow, by adding the rotation parameter $a$ the gas pressure and the density  reduce
 but the infall radial velocity and the angular velocity grow.
Also,  our conclusions indicate that stagnation
point, which separate the inflow and outflow regions, approaches  the central object by increasing the rotation parameter $a$ in co-rotating flow. However, the effect of $a$ on the behavior of the stagnation
point is inverted for counter-rotating flow i.e, this point  moves outward for a larger value of $a$.
This treatment suggests a way to distinguish between the co and
counter rotating accretion flow onto compact objects.

A unique aspect of advection dominated accretion flows is that they are characterised by positive values of Bernoulli parameter (NY94).
The  positive Bernoulli parameter shows that the accreting gas may be able to generate wind or outflow.
For a fully advection dominated flow ($f=1$) our solutions indicated that the entire accreting gas has a positive Bernoulli parameter.
Moreover, we showed that the Bernoulli parameter becomes more positive when  the effect of
rotation on the structure of disc decreases.

Here,  ratio of the intrinsic angular velocity of the central body to the rotating velocity of disc is assumed to be a constant value.
However,  the accreting particles will affect the evolution of the central object and will increase  angular momentum of it.
In a more realistic picture, this ration should be taken into account as a function of time and position.
Also, we ignored the latitudinal dependence of physical quantities,
although some authors have shown that latitudinal dependence is
important for the structure of a disc (Narayan \& Yi 1995; Ghanbari
et al. 2007, Habibi et al 2015). The latitudinal behaviour of such discs can be investigated in other studies.
 Furthermore,  we considered the advection parameter
$f$ be a constant. In a realistic model, this parameter can be
a function of position and time ; other researchers can take this
into account.




\end{document}